\documentclass[]{aa}
\usepackage[colorlinks=true,citecolor=blue]{hyperref}% To add links in your PDF file, use the package "hyperref"
\bibpunct{(}{)}{;}{a}{}{,} 
\usepackage[varg]{txfonts}
\usepackage{graphicx}   % Including figure files

\def\oldbf{}
\begin{document}

\title{Rotation periods and shape asphericity in asteroid families based on TESS S1-S13 observations}

\titlerunning{Asteroid families with TESS} 

\author{Gyula M. Szab\'o\inst{1,2,3,4}
        \and 
        Andr\'as P\'al\inst{3}
        \and
        L\'aszl\'o Szigeti\inst{1}
        \and
        Zs\'ofia Bogn\'ar\inst{3,5}
        \and
        Attila B\'odi\inst{3,5,6}
        \and
        Csilla Kalup\inst{3,7}
        \and
        Zolt\'an J. J\"ager\inst{1,2}
        \and
        L\'aszl\'o L. Kiss\inst{3,6,8}
        \and
        Csaba Kiss\inst{3,6}
        \and
        J\'ozsef Kov\'acs \inst{1,2,4}
        \and
        G\'abor Marton\inst{3,6}
        \and
        L\'aszl\'o Moln\'ar\inst{3,5,6}
        \and
        Emese Plachy\inst{3,5,6}
        \and
        Kriszti\'an S\'arneczky\inst{3,6}
        \and
        R\'obert Szak\'ats\inst{3}
        \and
        R\'obert Szab\'o\inst{3,5,6}}

\institute{ELTE E\"otv\"os Lor\'and University, Gothard Astrophysical Observatory, Szombathely, Hungary, %1
\email{szgy@gothard.hu}
\and
MTA-ELTE Exoplanet Research Group, 9700 Szombathely, Szent Imre h. u. 112, Hungary %2
\and
Konkoly Observatory, Research Centre for Astronomy and Earth Sciences, E\"otv\"os Lor\'and Research Network (ELKH), Konkoly Thege Mikl\'os \'ut 15-17, H-1121 Budapest, Hungary %3
\and
MTA-ELTE Lend{\"u}let Milky Way Research Group, Hungary %4
\and
MTA CSFK Lend\"ulet Near-Field Cosmology Research Group %5
\and
ELTE E\"otv\"os Lor\'and University, Institute of Physics, 1117, P\'azm\'any P\'eter s\'et\'any 1/A, Budapest, Hungary %6
\and
ELTE E\"otv\"os Lor\'and University, Department of Astronomy, 1117, P\'azm\'any P\'eter s\'et\'any 1/A, Budapest, Hungary %7
\and
Sydney Institute for Astronomy, School of Physics A29, University of Sydney, NSW 2006, Australia %8
}

\date{Accepted XXX. Received YYY; in original form ZZZ}

\abstract{
 Here we present the analysis of the distribution of rotation periods and light curve amplitudes {\oldbf based on 2859 family asteroids} in {\oldbf 16} Main Belt families based on 9912 {\oldbf TESS} asteroid light curves in the TSSYS-DR1 asteroid light curve database. We found that the distribution of the light curve properties follow a family-specific character in {\oldbf some asteroid families, including} the Hungaria, Maria, Juno, Eos, {\oldbf Eucharis, and Alauda} families.\ While in other large families, these distributions are in general very similar to each other. We confirm that older families tend to contain a larger fraction of {\oldbf more spheroidal}, low-amplitude asteroids. 
 We found that rotation period distributions are different in the cores and outskirts of the Flora and Maria families, while the Vesta, Eos, and Eunomia families lack this feature. We also confirm that very fast spinning asteroids are close to spherical (or spinning top shapes), and minor planets rotating slower than $\approx$11 hour are also more spherical than asteroids in the 4--8 hour period range and this group is expected to contain the most elongated bodies.}

%% !!!! GYULA EZT ELLENŐRIZD, HOGY JÓ-E!
\maketitle
%\label{firstpage}
%\pagerange{\pageref{firstpage}--\pageref{lastpage}}

\section{Introduction}

About one-third of all known asteroids belong to families \citep{1995Icar..116..291Z}, and clusters of asteroids are believed to originate from the collisional disruption of parent bodies \citep{2005Icar..178..179O}. Members of asteroid families share similar orbital elements \citep[semi-major axis, eccentricity, and inclination, see][]{2006IAUS..229..289N} and colours \citep{2002AJ....124.2943I}. The size distribution varies significantly among families, and it is typically different from the size distribution of background populations \citep{2008Icar..198..138P}. Often the slope follows a double power law, characterised by the two slopes and the size of transition between the two branches, and these three parameters vary among the families. The two branches of the double power law have a more dissimilar slope in the case of old families and families with S-type taxonomy \citep[see also][]{2008Icar..198..138P}.

\begin{figure}
    \begin{center}
    \includegraphics[width=\columnwidth]{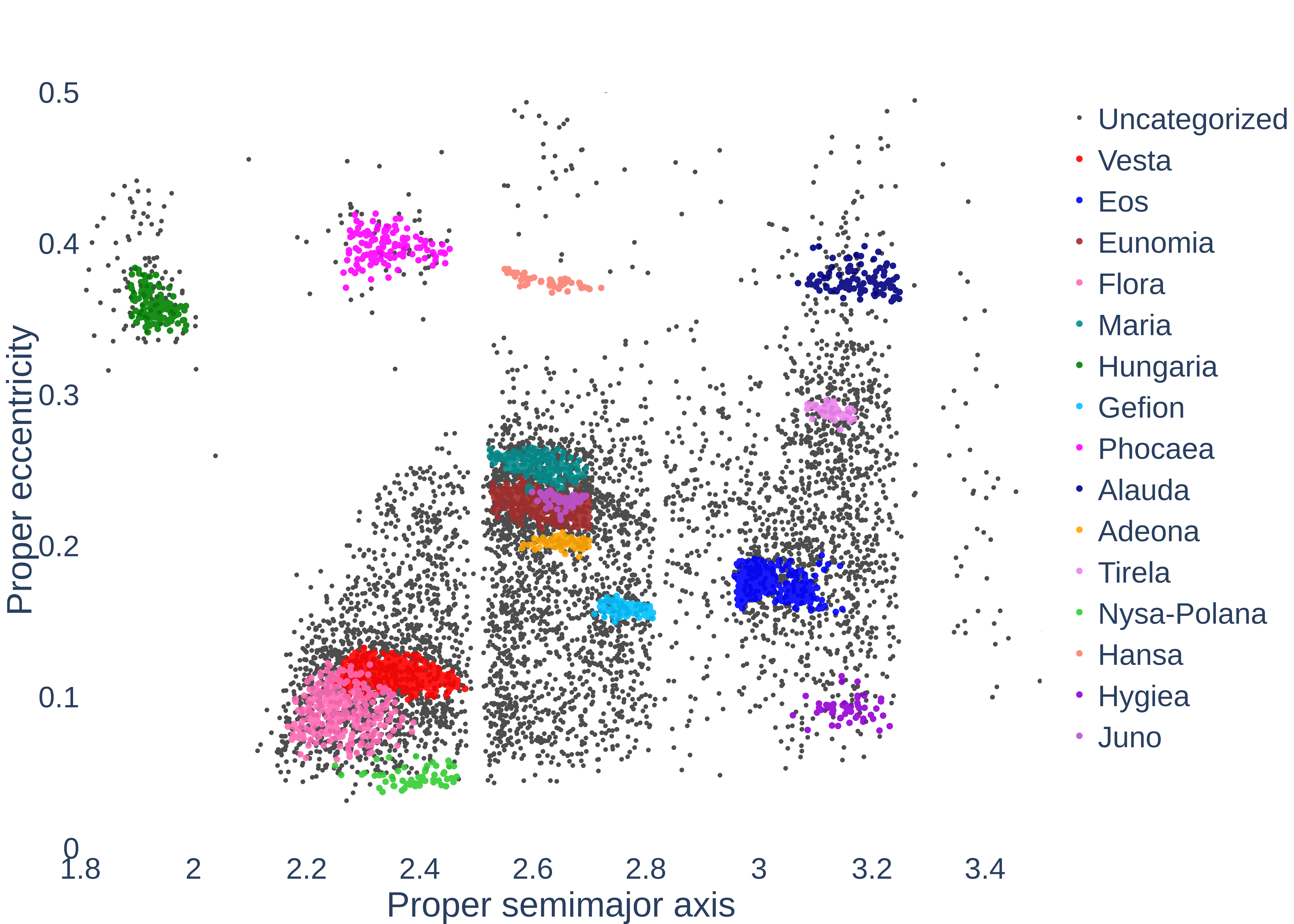} \hskip8mm
    \includegraphics[width=\columnwidth]{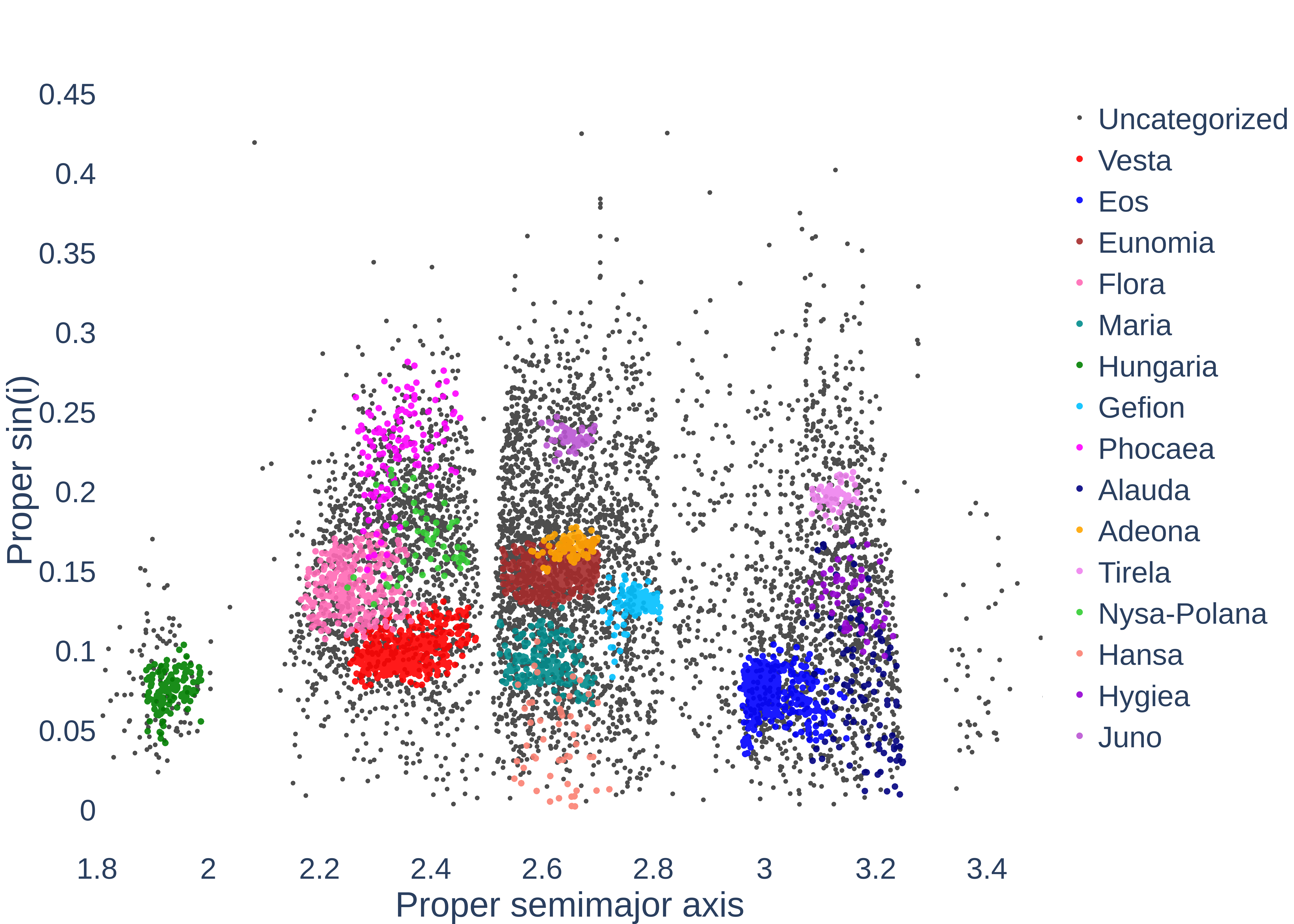}
    \end{center}
    \caption{TSSYS-DR1 asteroids in the orbital element space. Family members are marked by the colours as listed in the legend.}
    \label{fig:families}
\end{figure}

The rotation of family asteroids is widely accepted to reflect the collisional origin and the evolution of the asteroid family \citep[see e.g.][]{1992A&A...253..604F,2002aste.book..517P,2020NatAs...4...83C}. Rotation is also characteristic for a later evolution where internal collisions and YORP ({\oldbf Yarkovsky–O'Keefe–Radzievskii–Paddack effect};  \citealt{2006AREPS..34..157B,2015aste.book..509V}) are both acting. The collisions tend to evolve the rotation statistics of larger asteroids towards a Maxwellian distribution \citep{2002aste.book..113P}, which is the  boundary distribution of a collisionally relaxed state, while YORP \citep{2007Sci...316..272L,2015aste.book..509V} can increase the number of very slow rotators \citep{2020ApJS..247...26P} and also can spin asteroids up to reach the break-up barrier \citep{2014Icar..241...79P}.

The amplitude of the light curve reflects the minimum asphericity of the asteroids. The observed $A(\phi, \alpha)$ amplitude of light variation of a triaxial ellipsoid with axes $a>b>c$ is
\begin{equation}
    \frac{A(\phi,\alpha)}{1+m\alpha}=
     1.25\log\frac{(b/c)^2\cos^2\phi+\sin^2\phi}{(b/c)^2\cos^2\phi+(b/a)^2\sin^2\phi},
\end{equation}
where $\alpha$ is the solar phase, the $m$ parameter describes {\oldbf dependence of the brightness of the asteroid on the phase function}, and $\phi$ is the aspect angle of the rotational axis \citep{1986Icar...68....1M}. A non-orthogonal aspect compresses the amplitude. {\oldbf Expressing the amplitude in terms of} $\phi$ includes the pole coordinates as two implicit parameters that have to be fitted since the aspect angle depends on the mutual positions of the observer and the asteroid, as well as the direction of the rotational axis. When observations of several oppositions are known, all parameters can be fitted and the shape model can be reconstructed. From one single light curve, one cannot correct for $\phi$ and the observed amplitude is a lower estimate of the $a/b$ asphericity.

\cite{2008Icar..196..135S} followed the strategy above, but involving only changes to the brightness observed at two unrelated rotation phases close to opposition, as recorded in the SDSS Moving Object Catalogue \citep{2002AJ....124.1776J}. Since there were almost 12,000 asteroids in this analysis, the distribution of the asteroid asphericities could be determined and compared between families. The conclusion was that older families have less elongated asteroids on average, which was interpreted as evidence for a continuous slow shape evolution driven by impact-induced shaking events. As meteor impacts shake the asteroid body, the material slides towards the valleys of the actual shape which makes the asteroid shape 
%more rounded, or
closer to an oblate {\oldbf spheroidal ($a/b \approx 1$)} shape on a scale of a billion years \citep{2020NatCo..11.2655M}. The distribution of rotation periods was, however, not revealed by the sporadic observations in SDSS MOC.

In a recent study, \cite{2018A&A...611A..86C} compare the reconstructed distribution of shape $a/b$ axis ratios in different families with respect to that of background asteroids. They made use of Pan-STARRS1 asteroid detections and applied a clever reconstruction of the shape asphericity distribution function from sporadic precise detections. 
In accordance with \cite{2008Icar..196..135S}, they found that Massalia members are on average more elongated than their local background, and Phoacaeas members are even more elongated than those of Massalias. They did not conclude on systemic differences between asteroid families, but they recognised a particularly interesting feature: asteroids with rotation periods between 0--4\,h are more {\oldbf spheroidal} than slower rotators. \cite{2018A&A...611A..86C} also present a possible bimodal shape distribution in the case of the Phocaea family. 

In the first year of the {\it Transiting Exoplanet Survey Satellite}  \citep[TESS;][]{2015JATIS...1a4003R} mission, nearly ten thousand asteroid light curves were extracted with unambiguous determination of rotation characteristics \citep{2020ApJS..247...26P}. This data set enabled us to derive some fundamental physical properties for these objects and also to compare the distribution of rotation periods and amplitudes in different asteroid families. 

The outline of this present paper follows the aforementioned scope. In Section~\ref{sec:methods} we describe the selection of the asteroid families in this analysis, and a separation of the cores and outskirts. In Section~\ref{sec:results} we compare the period and amplitude distributions between families and also between the cores and outskirts in the
case of the most populated families. Section~\ref{sec:summary} gives a summary of our analysis.

\begin{table}
\caption{Asteroid families examined in this paper. All ages are taken from Bro{\v{z}} (2013), and references are given if there have been other investigations on the family age. Notes: $^1$ considered as old; $^2$ from cratering statistics.}
    \label{tab:families-ages}
    \begin{tabular}{lrcl}
Family  & No. & age      &      Further references\\    
        & &  (Myr)  & \\
\hline
Adeona  &  92 & 800     $\pm$   100     &        \\ %https://arxiv.org/pdf/1607.01998.pdf
Alauda  &  98 & $<$3500         &               \cite{2007DPS....39.1608M}$^1$\\
Gefion & 130 &  480     $\pm$   50&             \\
Eos     & 479 & 1300$\pm$               200     &       \\
Eucharis        &  58 & $<3500$  &
\\
%Euphrosyne &  47 &     {\oldbf 280}    & 100   & \cite{2004MPBu...31..100F}$^2$\\ %\cite{2020A&A...643A..38Y} \cite{2004MPBu...31..100F}$^2$
Eunomia & 450 & 1400    $\pm$   200     &       \cite{2014Icar..239...46M}\\
Flora   & 331 & 1000    $\pm$   500     &       \\
Hansa   &  51 & $<$1600 &       \\
Hungaria& 139 & 500     $\pm$   200     &       %Warner et al. (2010)
\\
Hygiea  &  51 & {\oldbf 3200} $^{+380}_{-120}$& \cite{2014MNRAS.437.2279C}              \\
Juno    &  51 & $<$700&         \\
Maria   & 184 & 3000$\pm$               1000&           \\
Nysa-Polana&  54 &      $\sim${\oldbf 2000}&            \cite{2013Icar..225..283W}      \\
Phocaea & 120 & {\oldbf 1200}   $\pm$   120     &       \cite{2009MNRAS.398.1512C}\\ %https://arxiv.org/pdf/1607.01998.pdf
Tirela  &  68 & $<$1000 &               \\
Vesta   & 503 & 1000    $\pm$   250     &       \\
\hline
    \end{tabular}
\end{table}

\section{The database}

Potential biases have been recognised to be present in ground-based wide-field surveys by  \cite{2011Icar..216..610W}. In that paper, a 'dense' data-set criterion was suggested describing the minimum level of sampling to avoid such biases:
eight non-clustered observations per night and eight nights per data set with a maximum time span of 30 days. We adopted this definition  here, and used only those TESS light curves in this study that fulfilled this criterion.

Rotation periods were determined by an algorithm that calculates the dispersion of the residual light curve for each frequency step, in a way similar to phase dispersion minimisation methods \citep{2020ApJS..247...26P}. All of these TESS light curves were checked manually for the appropriateness of the period found by the algorithms, the overall quality of the composite light curves, the lack of evident biases, satisfying phase coverage and the lack of cadence issues, and any other possible clues as to an inappropriate solution. All flagged cases then were reevaluated manually, and only those were used in our analysis where the general appearance was fully compatible with a proper light curve solution. In most of the cases, simple bimodal solutions with two maxima and minima per cycle were preferred, corresponding to elongated bodies, but more complicated light curves with a higher number of extrema per cycle were also identified.

Since the light curve shapes can be very diverse and Fourier methods are not always well-suited to determine the full amplitude of the variation, peak-to-peak amplitudes were determined as the brightness difference between the brightest and faintest phase bins.\ This is a standard method in asteroid photometry works. 

We also identified a small number of asteroids for which short periods below 2\,h were detected, and which would fall below the spin barrier for main-belt asteroids. Upon revision of these candidates, we found all of these to be classified as single-peaked (or multiperiodic, with a fast single-peak component) light curves. We assume that these objects actually have double-peaked light curves with perfectly symmetric humps (up to the level of the precision of TESS), and they are therefore slower-rotating objects in the 3--4~hr range, above the spin barrier. For this work we simply removed light curves with $<$2\,h periods. This cut affected 70 out of the 9912 object, or 0.7 per cent of our sample.

{\oldbf We also note that the spin barrier of 2h corresponds to spherical objects with a high density (S/M-type). Elongated C types (low density) have a disruption limit >2h, which could be even 3--4h. Therefore, it is possible that some of the periods in the 2--3h range could still correspond to single-peaked light curves, thus values of P/2.}

\section{Methods}
\label{sec:methods}
Asteroid families were taken from the literature. The family membership flags in the TESS Asteroid Light Curve Catalogue TSSYS-DR1 are {\oldbf taken from the Asteroid Family Catalogue} \citep{2015aste.book..297N}.
{\oldbf We did not classify or reclassify the asteroid family memberships; the flags from TSSYS-DR1 were taken as categorical variables in the forthcoming analysis. The family assignments of TSSYS-DR1 asteroids are shown in Fig.~\ref{fig:families}. }

The most populated {\bf five} families {\bf (Vesta, Eos, Eunomia, Flora, and Maria)} were involved in a comparative analysis of period and amplitude distributions {\oldbf between the core and the outskirts of these families. The working hypothesis was that an internal structure of the asteroid families can be observed in the rotation parameter space just as it has been confirmed for the size distributions \citep[e.g.][]{2008Icar..198..138P}. This could be a result of some internal differentiation process which is sensitive to the rotational state, or simply due to contamination of background asteroids that were scattered in the family. Both processes can lead to a different amplitude and period distribution between the cores and the outskirts. 

The family members were}
separated by their distance to the centre. Due to correlations of parameters, the asteroid families are more or less elongated in the space of orbital parameters. Therefore, we used the Mahalanobis distance \citep{mahalanobis1936generalized}, where the distance metric is derived from the covariance matrix of the sample, {\oldbf and such way, the elongation of the distribution is handled properly.}

The Mahalanobis distance of an asteroid with orbital elements $\vec{o}= (a, e, \sin{} i)$ ) from a set of asteroids in the set $\mathbb{O}=\{ \vec {o}_1,\vec {o}_2... \} $  is defined as follows: 
\begin{equation}
D_{M}({\vec {o}, \mathbb{O}}):={\sqrt {(\vec {o}-\langle \vec {o}_i\rangle
)^{T}\ S^{-1}\ (\vec {o}-\langle \vec {o}_i\rangle)}} \, ,
\end{equation}
where $S$ is the covariance matrix of the orbital elements, $\langle \rangle$ denotes the average, and $\vec {o}_i$ represents all the members of the $\mathbb{O}$ asteroid family here. 

This method normalises the standard deviation of the distribution of the family asteroids along the primary components in the orbital space, and it measures the distance in this geometry. We defined the outskirts of a family when $D_{M}(\vec{o}, \mathbb{O})>
Q(D_{M}(\vec {o}_i, \mathbb{O}),0.67)$, that is the outermost 33 percentile of the family, where $Q(,0.67)$ means the 67\%{} quantile. The `inner' 67\%{} of family members are considered as the core of the family.

{\oldbf
This border between the core and the outskirt seems to be somewhat arbitrary, but it is necessary to support enough asteroids in both categories. There are no astrophysical considerations behind this splitting, just mathematical criteria to avoid low-number statistics in any of the categories.
According to our methodology, the distribution family members were mapped to a spherically symmetric distribution (via the Mahalanobis distance), which is very close to multivariate normal distribution. The criterion of 67\% corresponds to separating the `core' and the members that are at least 1 $\sigma$ distance from the centre of the cluster --  we refer to this as `outskirts' hereafter.}

Family ages were collected from the literature, with most age entries taken from \cite{2013A&A...551A.117B} (see the ages and their references in Table~\ref{tab:families-ages}). For most families, an age estimate was available, but some families had only upper limits or lacked an age estimate.

{\oldbf
The Hygiea family is listed to be at least 2$\pm$1 Gyr old in \cite{2013A&A...551A.117B}, while \cite{2014MNRAS.437.2279C} give an age estimate of
3200$^{+380}_{-120}$ Myr, placing this family among the oldest ones in our analysis.
%Euphrosyne family has been recognised to have formed via a reaccumulation process instead of a cratering event, and recent N-body simulations indicate that the age of the family is 280 $^{+180}_{-80}$ Myr \citep{2020A&A...643A..38Y}, which is well within the upper limit of 1.5 Gyr by \cite{2004MPBu...31..100F}. 
The dynamically defined complex asteroid family Nysa-Polana \citep{2001Icar..152..225C} has recently been recognised to incorporate an older and more widespread primitive family in the same region of the asteroid belt parented by asteroid (142) Polana, which was formed more than 2000 Myr ago \citep{2013Icar..225..283W}, whose complex has been referred to as the Eulalia and new Polana families more recently. Here we keep the older definition of this complex because it has only 54 entries in the TSSYS-DR1. Splitting the complex into two families would lead to asteroid counts below 50 and to the omission the both families within this complex from the entire analysis.

When comparing the distributions and testing whether they originate from different distributions, we applied the well-known KS test \citep[e.g.][]{1993stp..book.....L}, which compares the full shape of the distribution.\ We also applied the Wilcoxon test \citep[e.g.][]{1993stp..book.....L}, which is a non-parametric rank test to compare the median of the distributions. 
}

\section{Results}
\label{sec:results}

\subsection{Fast and slow rotators}

\begin{figure*}
    \centering
    \hbox{
    \includegraphics[bb=8 55 473 404, width=0.5\textwidth]{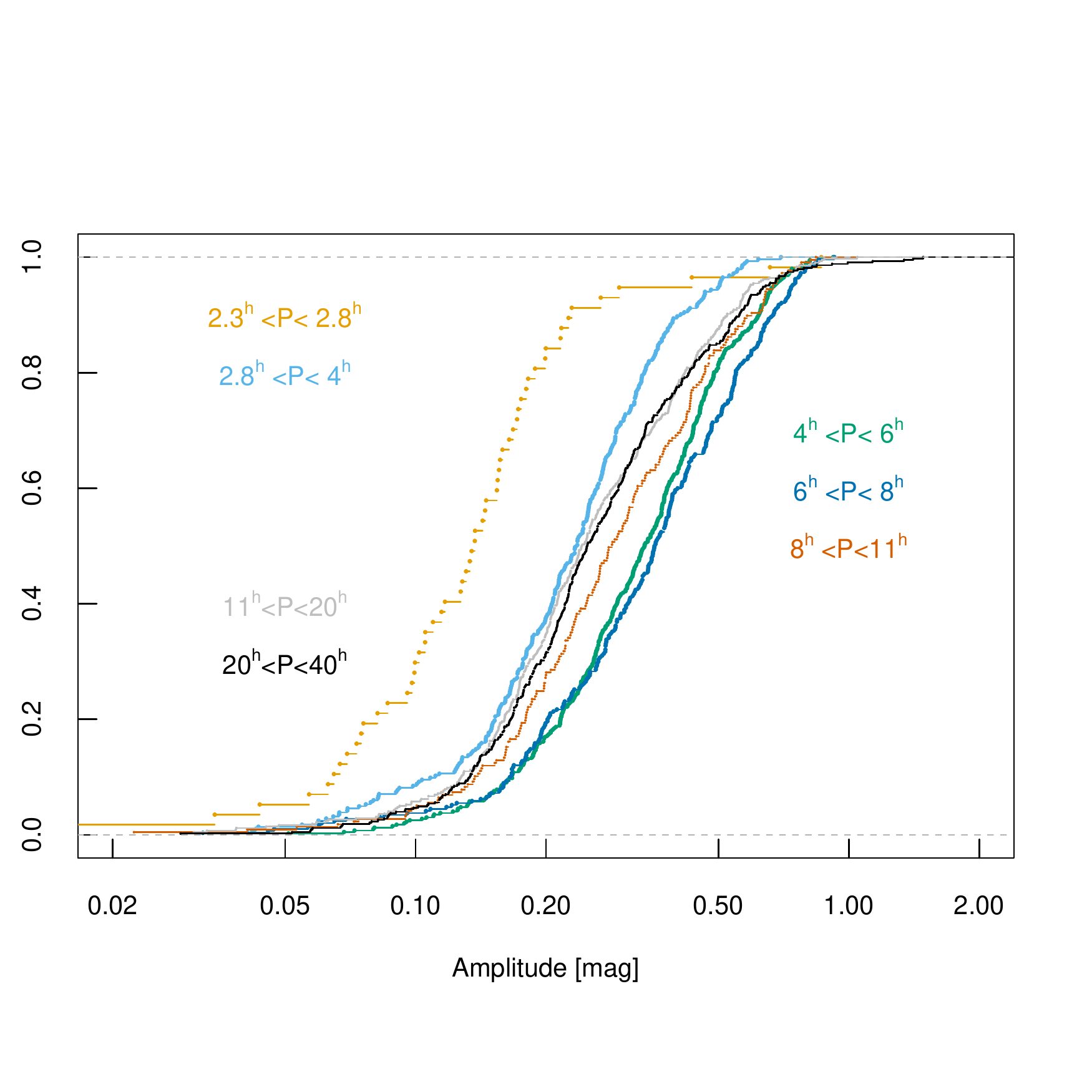}
    \includegraphics[bb=8 55 473 404, width=0.5\textwidth]{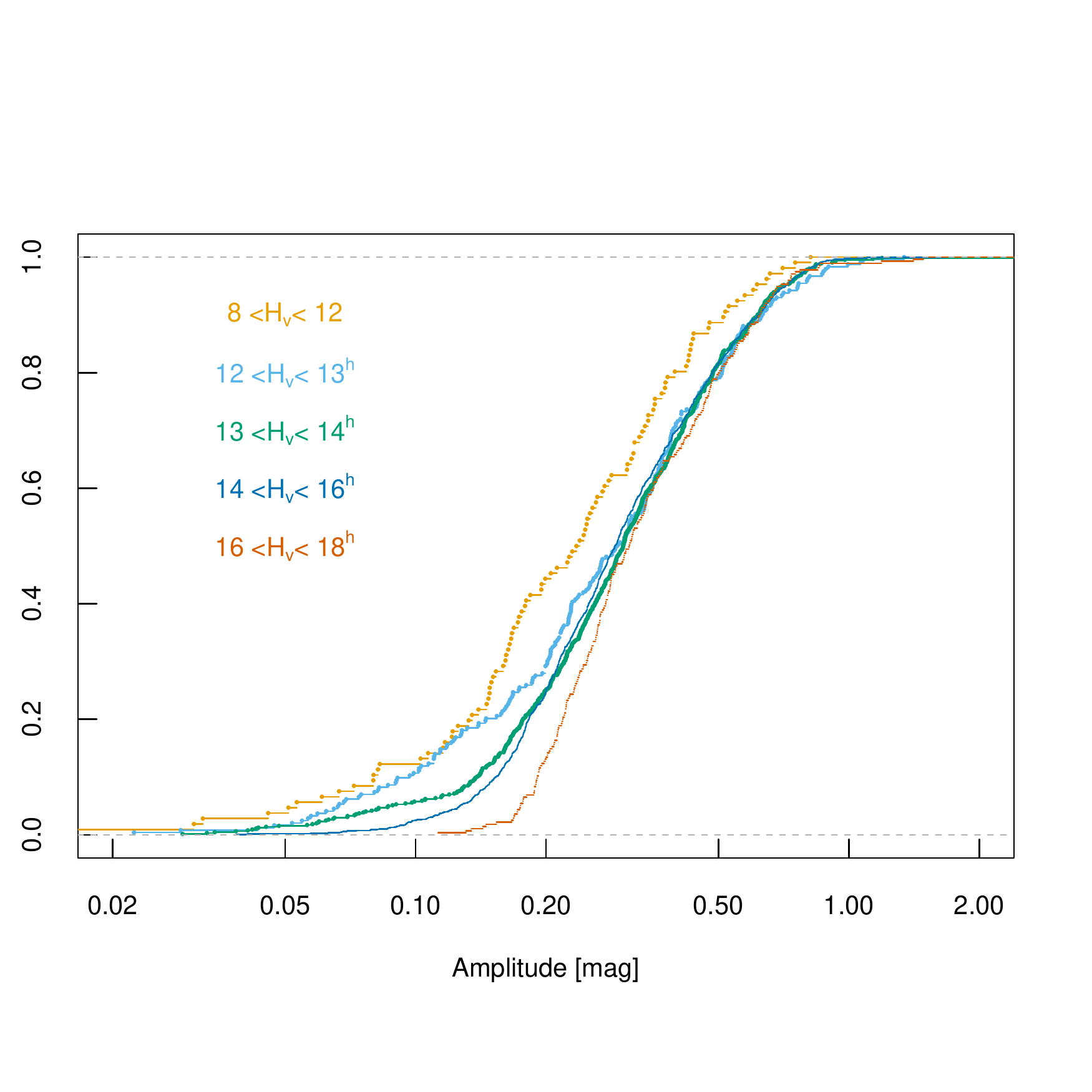}}
    \caption{Light curve amplitude distributions of family asteroids. Different curves represent asteroids from all families in various ranges of a rotational period (left panel) and $H_v$ absolute brightness (right panel).}
    \label{fig:fsrotators}
\end{figure*}

\cite{2018A&A...611A..86C} show that fast rotating asteroids have more {\oldbf spheroidal} shapes as they are characterised by an excess of $b/a\approx 1$ in the axis ratio distribution. They compare two groups of asteroids, split at a 4\,h rotation period.

We repeated this analysis with a larger number of asteroids in our sample. The results are presented in Fig. \ref{fig:fsrotators}. The fast rotating asteroids were indeed found to have smaller light curve amplitudes, indicating their more {\oldbf spheroidal} shape. This is most prominent in the case of asteroids with {\oldbf a 2}--3~h rotation period which share a median light curve amplitude of 0\fm15. The 3--4~h period group is far from being this extreme, although it can also be characterised by the lack of asteroids exceeding amplitudes of 0\fm45--0\fm5. The largest $b/a$ values are expected in the 4--8~h rotation period range where less than 10\% of asteroids were observed with amplitudes smaller than 0\fm15, with the median being in the 1--3\,h group.

\begin{figure*}
    \centering
    \hbox{
    \includegraphics[bb=50 90 394 470, width=0.48\textwidth]{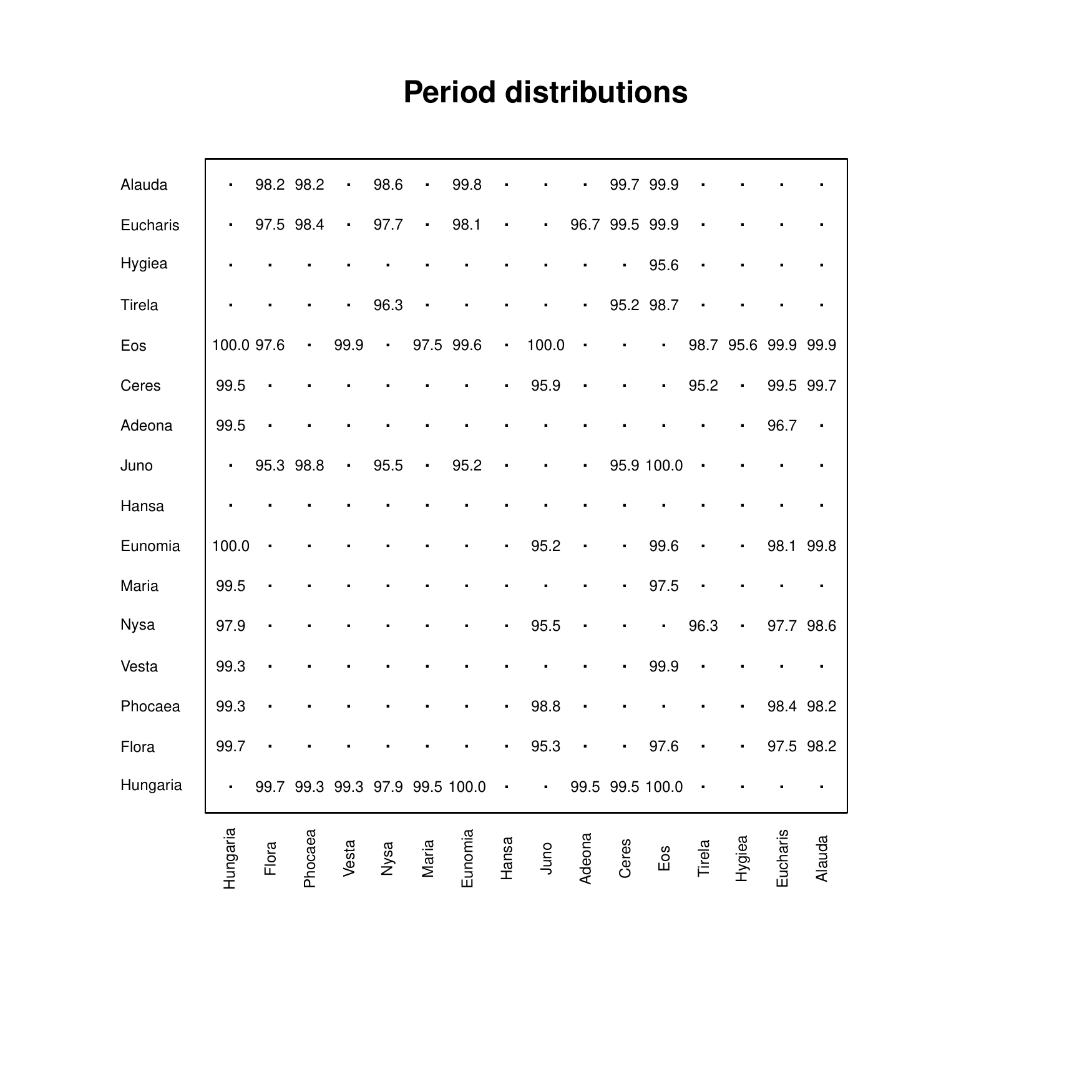} \hskip0.8cm
    \includegraphics[bb=50 90 394 470, width=0.48\textwidth]{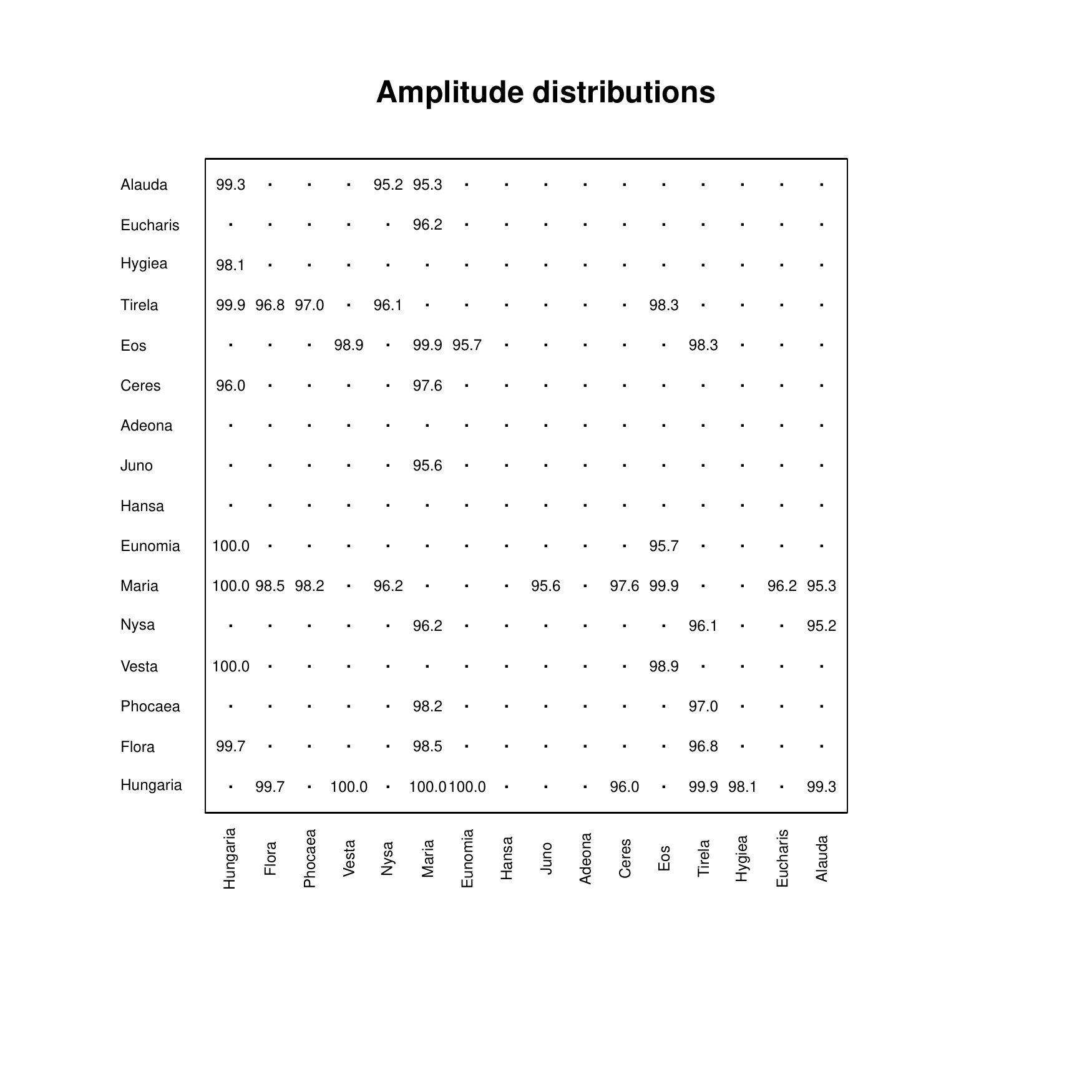}}
    \caption{Matrix of $1-p$ percentages of amplitude and period distributions in families, compared by KS tests. 
    {\oldbf The families are ordered by heliocentric distance.}}
    \label{fig:familyrotation}
\end{figure*}

Interestingly, the number of small amplitude asteroids starts to increase again with the increasing period, which is a completely unexpected feature. In the left panel of Fig.~\ref{fig:fsrotators}, we can see that the amplitude distributions of asteroids in the 11--20 and 20--40\,h period regimes are very similar and very close to the fast rotating asteroids in the 3--4~h group; they also share a median observed amplitude of 0\fm25. The only difference between the two groups is that 10\% of asteroids in the slow rotating group exhibit amplitudes $>0\fm5$, which is completely unobserved in the 3--4~h group.

The excess of rounded bodies among the fast rotators can be easily explained by the effect of the centrifugal force, {\oldbf which drives the shape closer to an ellipsoid on a long timescale as the rubble slides, which is due to micro-impacts, thermal erosion, weathering, etc.} The puzzling part is the case of slow rotators. It is hard to invoke any single process here. For example, the YORP effect has symmetry in the case of spheres and acts complicatedly on elongated bodies, so one can suggest that there is a selection  effect here: the more elongated bodies simply spin up in a short timescale and they are not observed as slow rotators any more. However, this does not explain the large-amplitude slow rotators. Slow rotation is also considered as a marker of binarity, and it may be that in this group the light curve is dominated by the orbital motion of the components, similar to binary stars, and the outcome is a slow variation with a large amplitude. Also, there can be contact binary asteroids among the large amplitude slow rotators.

It is also known that larger asteroids tend to be closer to a spherically symmetric shape than smaller asteroids. This means that smaller asteroids are over-represented in the groups of fast rotators in the comparison that we have just discussed. How this selection effect distorts the result can be seen in the right panel of Fig. \ref{fig:fsrotators}. As expected, we see a significant dependence here, which is less prominent as in the case of the period, and with the decreasing size (increasing $H_v$), the variation of the curves is monotonic. The most prominent differences are seen for the largest asteroids in the group of $8<H_v<12$. (We must note that very few asteroids fall in the 8--10 range, and this group mostly represents asteroids in the $H_v$ range of 10--12.) In this group, the entire curve is above the other curves, meaning that the amplitude distribution is biased towards the smaller amplitudes in all amplitude ranges. 

Asteroids in the 12--16 range of $H_v$ show a very similar amplitude distribution, and the amplitude distribution of the smallest asteroids ($16<H_v<18$) diverges only in the range of 0.1--0.3 magnitude towards larger amplitudes. All these curves are indistinguishable above an amplitude of 0.32, showing that the proportion of large amplitude asteroids (with amplitudes larger than 0\fm32) are the same for all asteroids in this sample. The under-representation of small amplitude faint asteroids (the lower part of the red curve) can also be a selection effect, at least in part: the light curves of fainter asteroids are more noisy, so their brightness variation can be hardly recognised in the noise and they suffer a more probable rejection when building the TSSYS-DR1 catalogue  up.

In essence, the size-amplitude dependence is less prominent than the period-amplitude dependence, and for asteroids with amplitudes exceeding 0\fm32 the dependence is even narrower and is only seen for the largest asteroids. Therefore the difference in the curves in the left panel of Fig.~\ref{fig:fsrotators} cannot be explained as a bias of a different size distribution in different sub-samples. Also, the non-monotonic nature of the period--amplitude distribution cannot be explained by the monotonically varying size distributions. The period--amplitude variation is therefore a complex phenomenon that probably has more divergent roots than the -- more simple -- brightness-amplitude variations.

\subsection{Period and amplitude distributions in families}

The derived amplitude and period distributions are shown for the examined {\oldbf 16} individual families in the Appendix. We compared the cumulative period and amplitude distributions in these families by applying a KS test and also a Wilcoxon test to the measured rotation parameters in all possible family pairs. The simplified test matrix {\oldbf ordered by heliocentric distance} is summarised in Fig.~\ref{fig:familyrotation}. We plotted a $1-p$ value expressed in per cent for the statistics when $p<0.05$, otherwise the field was left blank. We note that $1-p$ values express the expectation probability of two realisations from the unified distributions being less dissimilar than the observed one; so if $1-p$ is very close to 100\%,{} it is very unlikely that the background distributions of the two compared families are the same. The families are ordered according to the increasing median semi-major axis.

Interestingly, the differences in the period and amplitude distributions follow different patterns. In the period distribution test matrix, significant values are mostly observed between families at a larger solar distance, and the Hungaria group seems to be different from many other families. On the contrary, the amplitude distributions are found to be different, mostly between families closer to the Sun, and more similar to each other at larger solar distances.

The test matrix of period distributions is significant in quite a few family pairs. While there are families that differ more from the others than the average (such as Eos, Juno, and Hungaria), there is no dominant outlier, and all families except for Hansa differ from at least one other asteroid family, in respect to the period distribution.

The test matrix of amplitudes shows a completely different structure. Hungaria and Maria families seem to be quite unique, their period distribution differs from most of the other families very significantly. The amplitude distribution of Tirela asteroids also differs from Hungaria and four other families. By excluding Maria and Hungaria, the pattern shows only subtle differences between the families, and there are seven families showing no differences. 

The two test matrices show quite weak similarity. For example, the most significant families in the amplitude tests behave very differently in the period test matrix. Hungaria members show differences from nine families in the period distribution, while Maria members differ from only two families. One of them is the Hungaria family, and the other is Eos, the most significant family in the period distribution statistics.

Explaining why the family-level structures of period and amplitude distributions are so different remains beyond the scope of the present paper. This comparison would also be useful to control the albedo (at least for those families that have a single characteristic albedo) and to understand the size--amplitude distributions with a better resolution. This, however, requires performing the analysis at the family level, with many families containing 500--1000 records in the database. This cannot be fulfilled based on the TSSYS-DR1 asteroid catalogue, but in its later issues there will be enough data to redraw Fig. \ref{fig:fsrotators} at the family level.

Although a family-to-family comparison is not possible, we can just concentrate on the example of Eos and Maria families -- with a very marked behavior in one test and an almost unnoticeable one in the other test -- or families that show less prominent, but a very similar behaviour (Flora, Vesta, Hungaria, etc.). They demonstrate that the amplitude distributions in an asteroid family can hardly be predicted from the period distribution and vice versa. Therefore we can conclude that the origin and formation of the present shapes and the present rotation rates evolved in a complex way and at least one of the dominant steps was eventually a different process. There are many points where the period and shape evolution can meet -- such as impacts, the YORP effect, centrifugal shaping of fast rotators, mergers of binaries, and torque from the plane of the Ecliptic -- while explaining the detected differences seems to be more challenging.

\subsection{Age dependence}

\begin{figure}
    \centering
    \includegraphics[viewport=42 70 442 370,
width=\columnwidth]{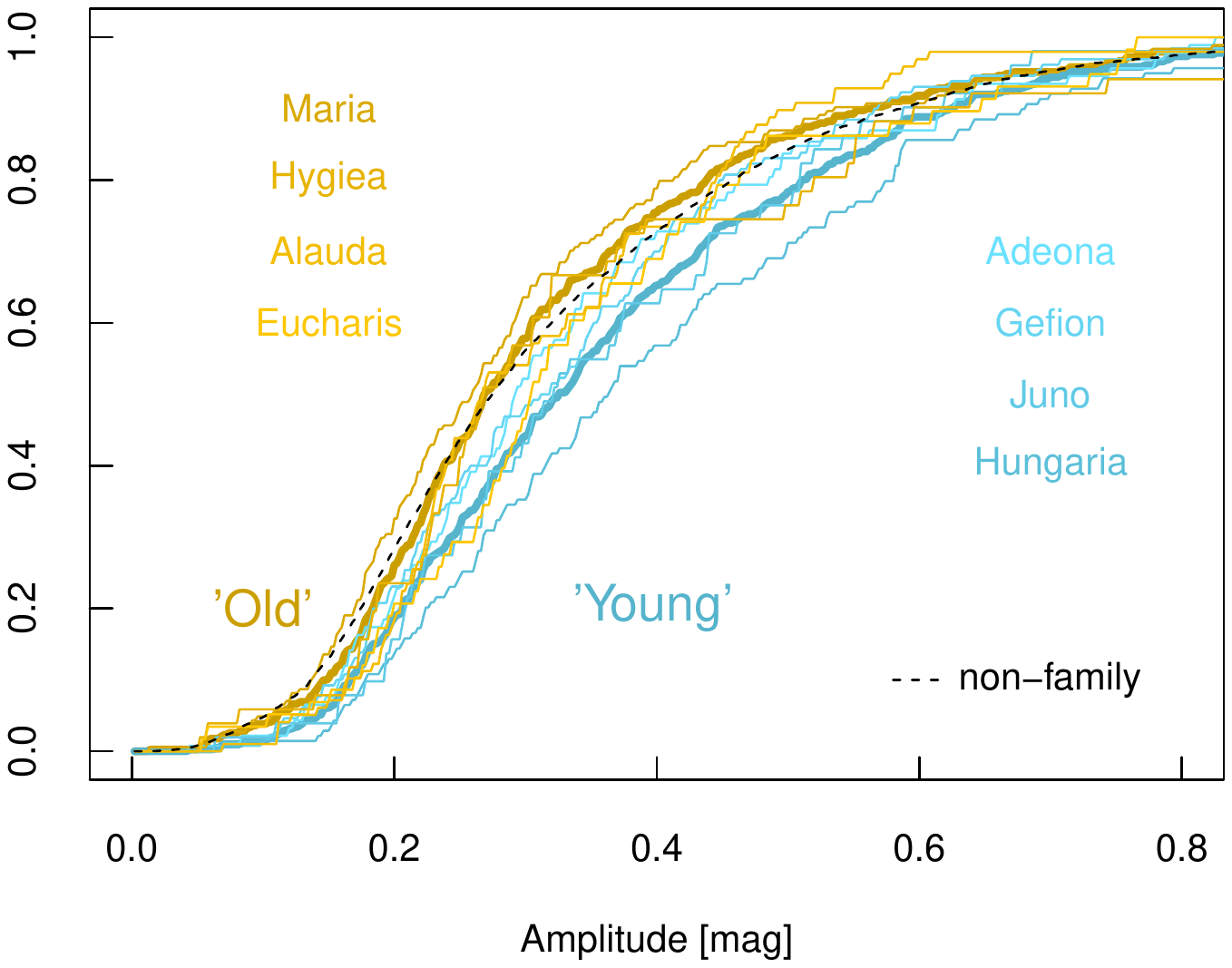}
    \caption{Amplitude distribution in four old families (thin brownish lines) and in four young families (thin blue lines). The lowermost blue curve is for the Hungaria family. The thick lines show the amplitude distribution in the young and old groups together. The black dashed line represents the non-family asteroids.}
    \label{fig:yo_dist}
\end{figure}

In Fig. \ref{fig:yo_dist} we compare {\oldbf four asteroid families with the youngest confirmed age in Table 1 (Adeona, Gefion, Hungaria, and Juno,} which are younger than 700 million years - thin blue lines) with four confirmed old families (Maria, Hygiea, {\oldbf Eucharis}, and Alauda, which are all older than 2 billion years - thin brown lines). We did not label the individual curves with the family labels which cross each other, and labelling would eventually lead to typographical confusion. However, this is not a problem since we have proven with KS and Wilcoxon tests that the amplitude distribution within the old families are indistinguishable from each other, and they represent an identical or very similar amplitude distribution. The amplitude distributions in the Adeona and Gefion families are also indistinguishable, while the Hungaria group represents an amplitude distribution even more skewed towards large amplitude variations (the lowest curve is in  Fig. \ref{fig:yo_dist}.)

Particularly meaningful are the amplitude distributions in 'combined old' and 'combined young' samples, plotted with thick brown and blue lines in Fig.~\ref{fig:yo_dist}. They are significantly different from each other: a KS test gives $p=0.0002$, while the Wilcoxon test results in $p=0.00005$. This is very significant, and it has very important implications on the evolution of asteroid shapes with the age of the family.

To compare the amplitude distributions in old and young families, the non-family asteroid sample is superimposed with a dashed line. Evidently, the amplitude distribution of non-family members and that of old families are indistinguishable, while the young families are notably different.

\citet{2008Icar..196..135S} found a similar age--asphericity relation in the sparsely sampled SDSS data. In young families, asteroids have a wide range of shape elongations, implying fragmentation-formation. Whereas in older families we detected an increasing number of rough spheroids, which is in agreement with the predictions of an impact-driven evolution. This was interpreted as observational evidence for the predicted systematic slipping down of the rubble on the surface towards the valleys and the sides of the body, caused by shaking due to the quite regular micro-impacts \citep{2000Icar..146..133L, 2004Sci...306.1526R}. 

In addition to  impact shaping, asteroids also suffer excavation processes due to micrometeoroid impacts. \citet{2009ApJ...699L..13D} found that if the excavation is dominant and the sliding of the material can be neglected, the shape evolves towards an increasingly elongated low-order polynomial, with an edgy tetrahedron being the final stage. Also \cite{2015MNRAS.454.1704H} predict an evolution towards elongated shapes if sliding material is neglected, although they remark that since the timescale of such impacts is much longer than the asteroid collisions, such evolution can barely be observed. 

The detection of the differences between asteroid shapes in different families in TSSYS-DR1 asteroid data confirms that the older families host more {\oldbf spheroidal} asteroids. This finding is more or less consistent with the predictions and observations so far, and it is a strong indication for shape evolution models in which seismic shaking and material redistribution due to sliding are the dominant processes.

\begin{figure}
    \centering
    \includegraphics[viewport=14 70 432 380,
width=\columnwidth]{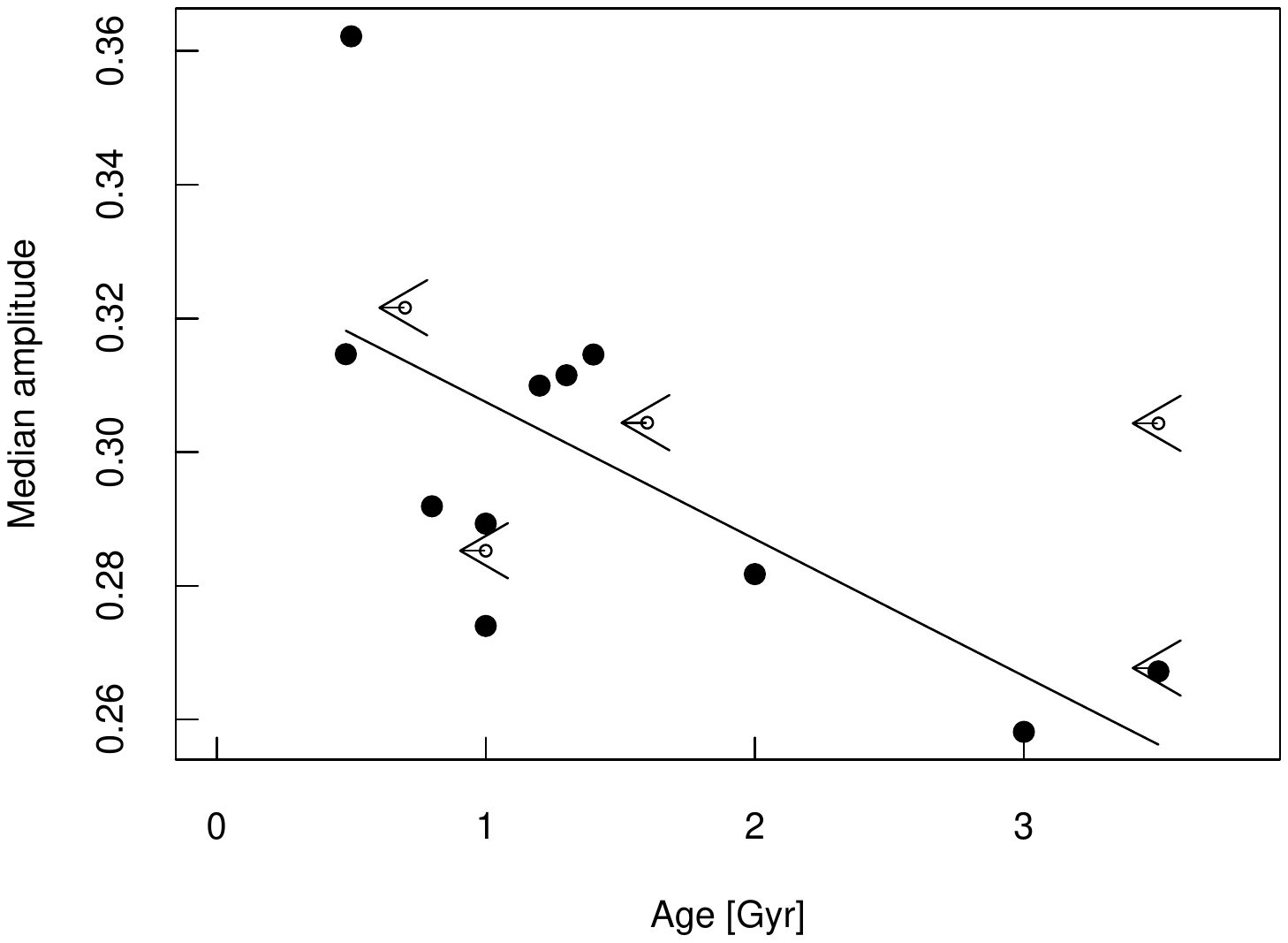}
    \caption{Median of the measured amplitude in asteroid families, as a function of their age. Families with an upper age limit only are plotted with open symbols and arrows indicating a possible younger age. The line is fitted to the filled circles.}
    \label{fig:yo_scatter}
\end{figure}
%%%%%

%\begin{figure*}[ht!]
%\centering
%\hbox{
%\includegraphics[bb=7 50 473 431,width=.33\textwidth]{florap_cc.eps}
%\includegraphics[bb=7 50 473 431,width=.33\textwidth]{mariap_cc.eps}
%\includegraphics[bb=7 50 473 431,width=.33\textwidth]{vestap_cc.eps}
%}
%\hbox{
%\includegraphics[bb=7 50 473 420,clip,width=.33\textwidth]{floraa_cc.eps}
%\includegraphics[bb=7 50 473 420,clip,width=.33\textwidth]{mariaa_cc.eps}
%\includegraphics[bb=7 50 473 420,clip,width=.33\textwidth]{vestaa_cc.eps}
%}
%\caption{Period and amplitude distributions in the cores and the outskirts of three families. }
%\label{fig:cc}
%\end{figure*}

\begin{figure*}[ht!]
\centering
\hbox{
\includegraphics[bb=7 50 473 431,width=.33\textwidth]{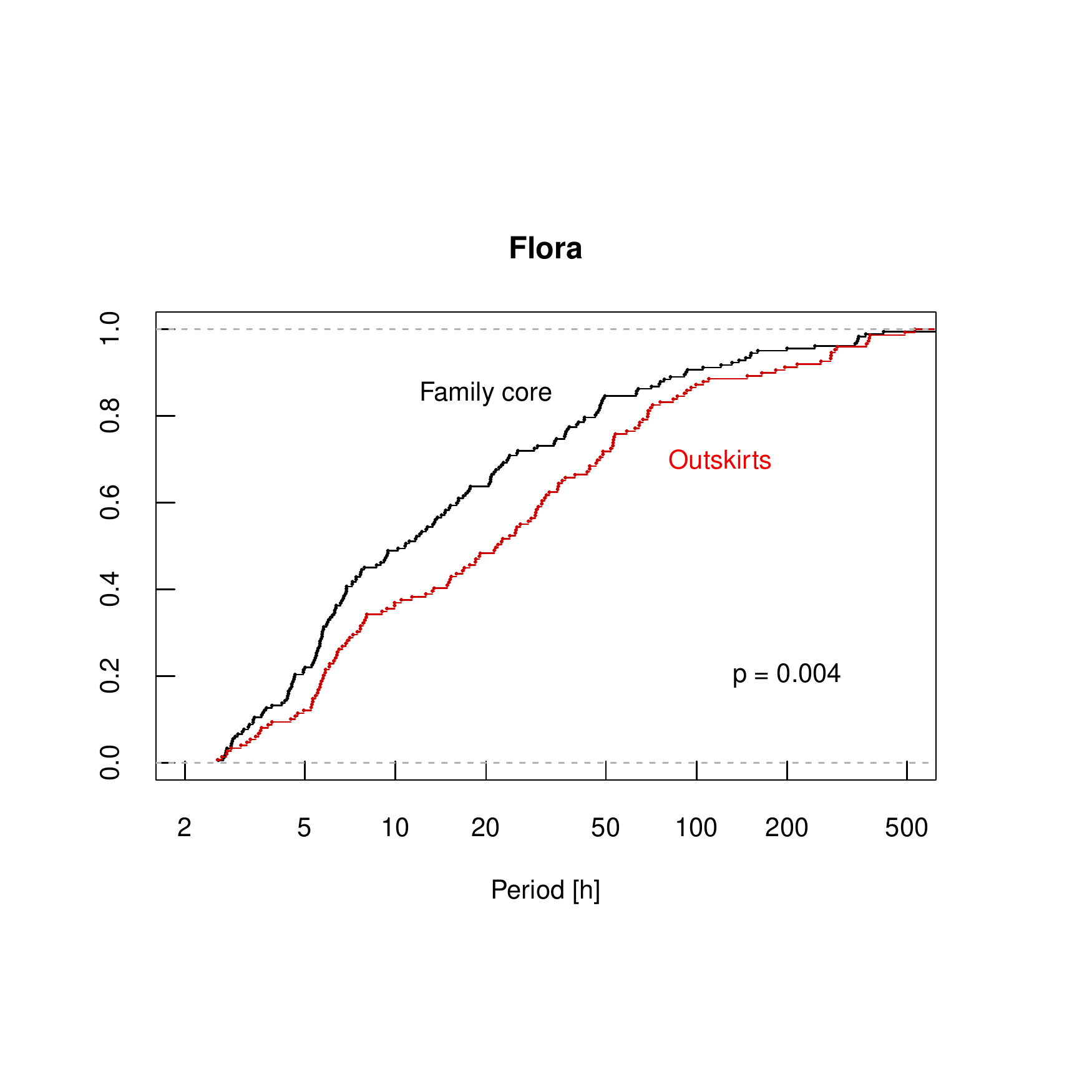}
\includegraphics[bb=7 50 473 431,width=.33\textwidth]{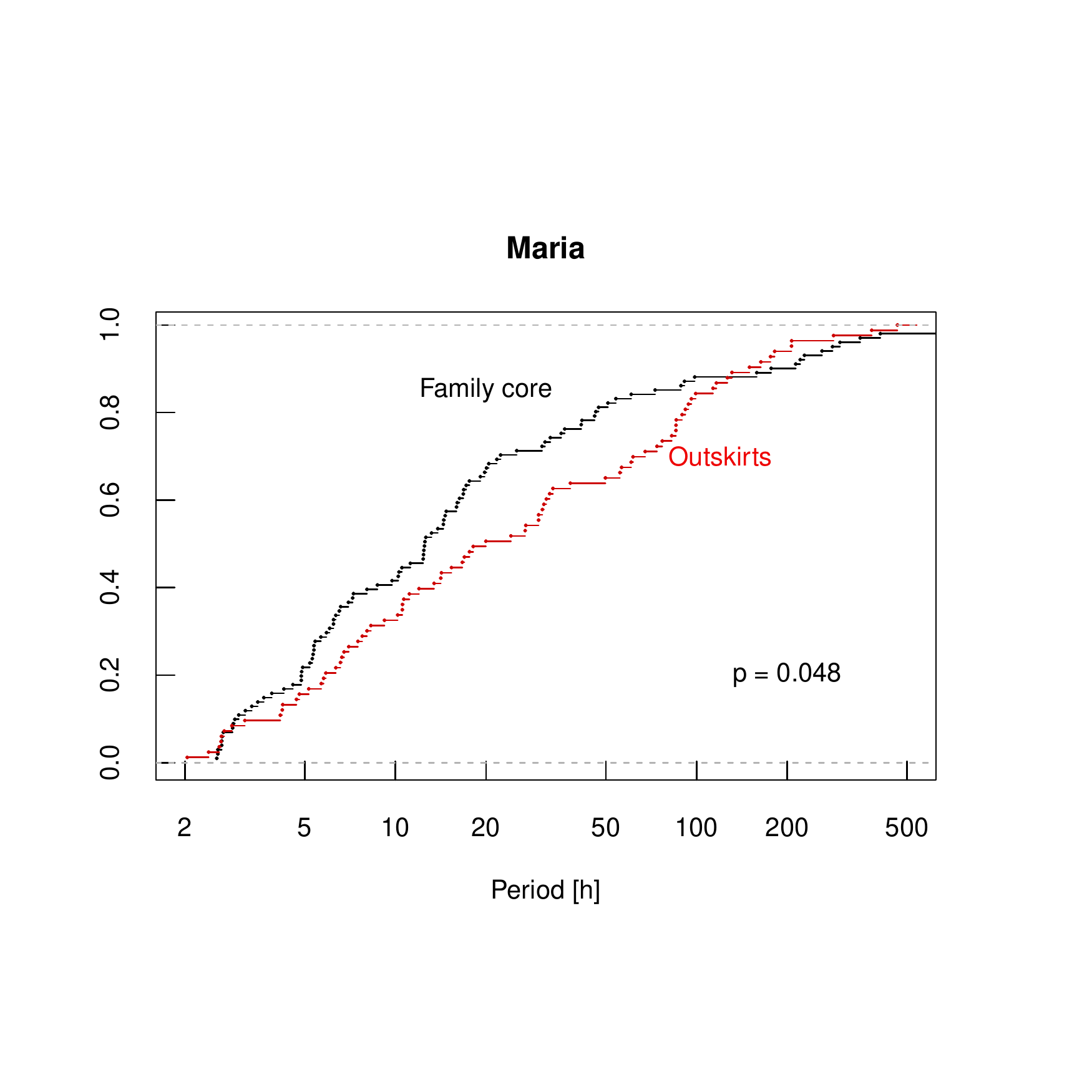}
\includegraphics[bb=7 50 473 431,width=.33\textwidth]{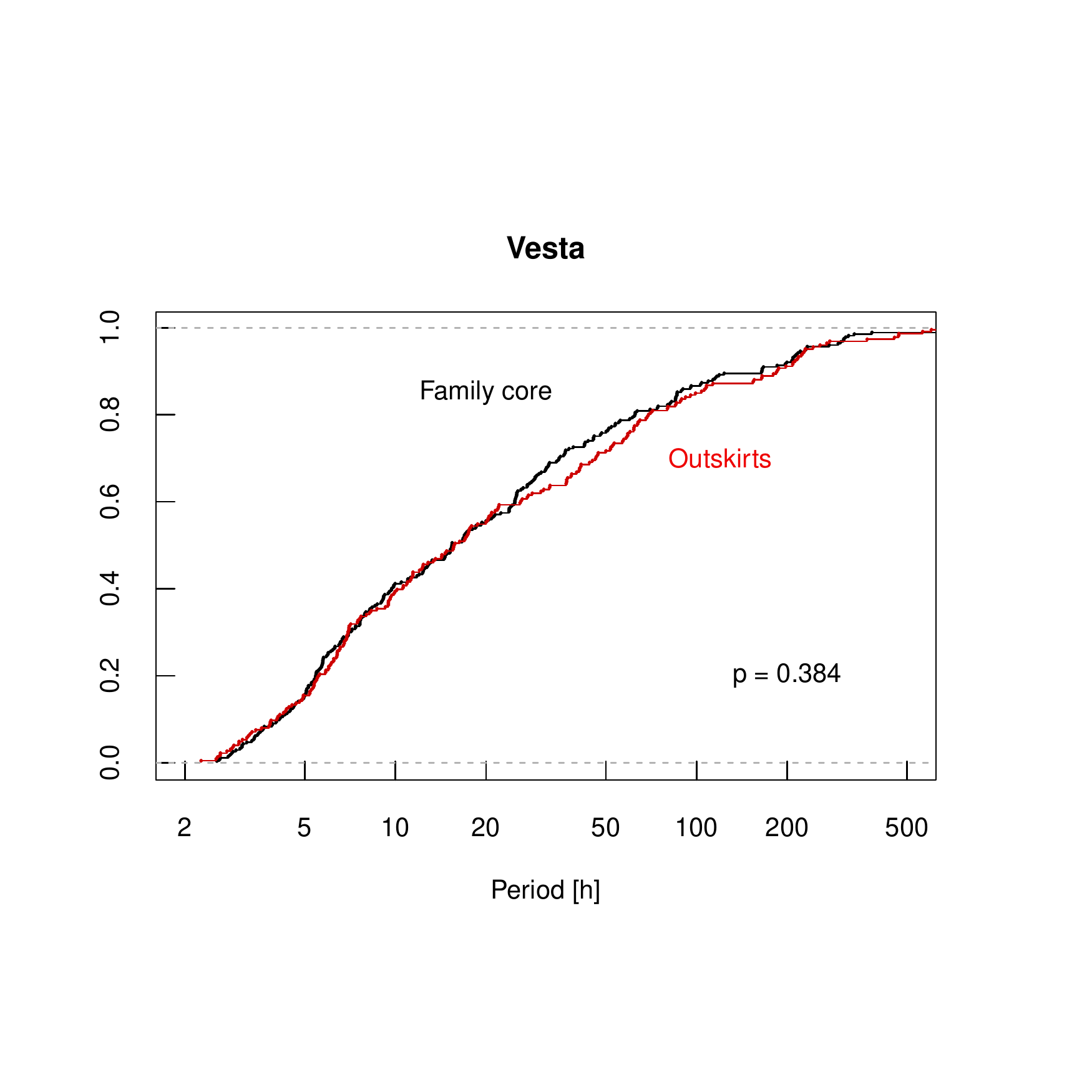}
}
\hbox{
\includegraphics[bb=7 50 473 420,clip,width=.33\textwidth]{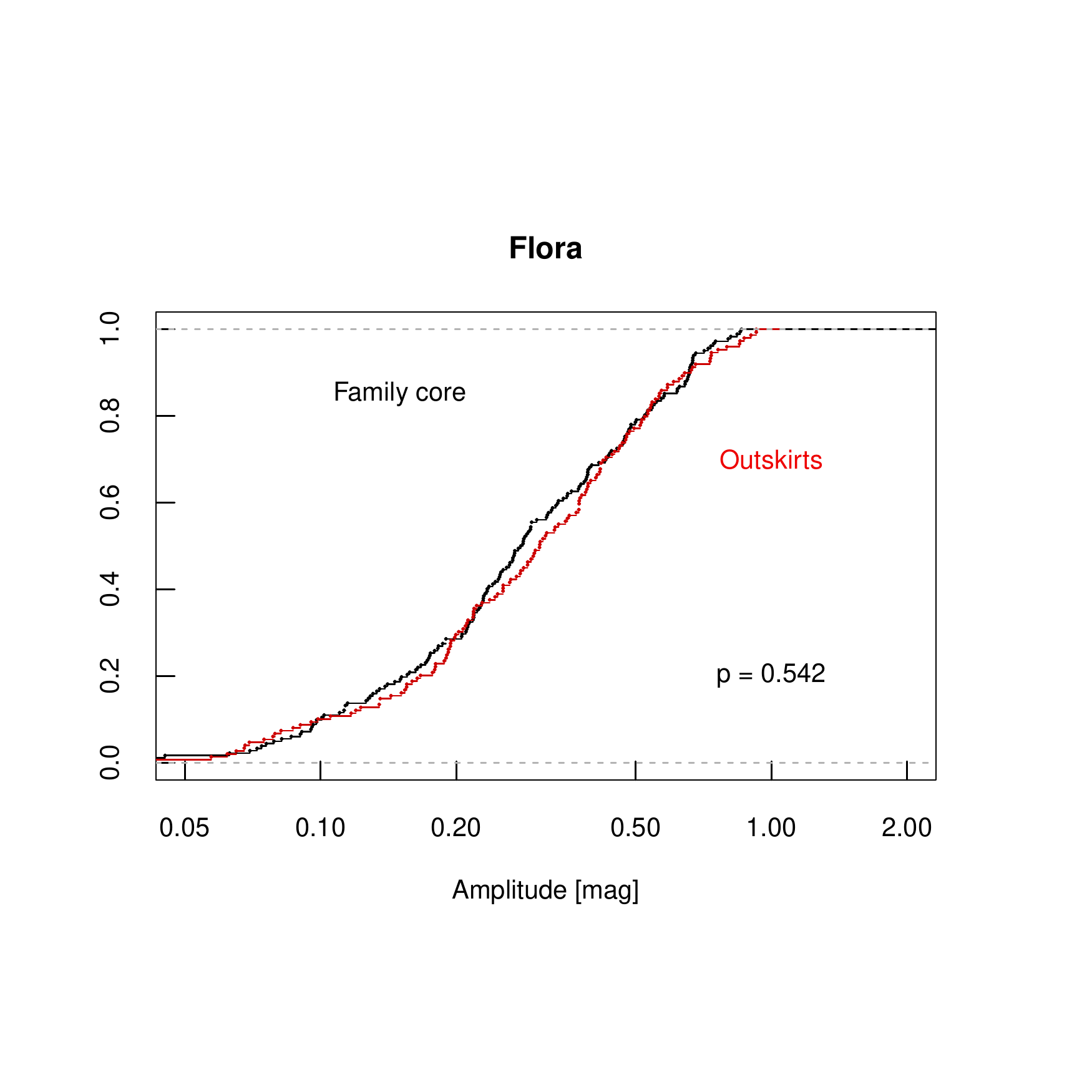}
\includegraphics[bb=7 50 473 420,clip,width=.33\textwidth]{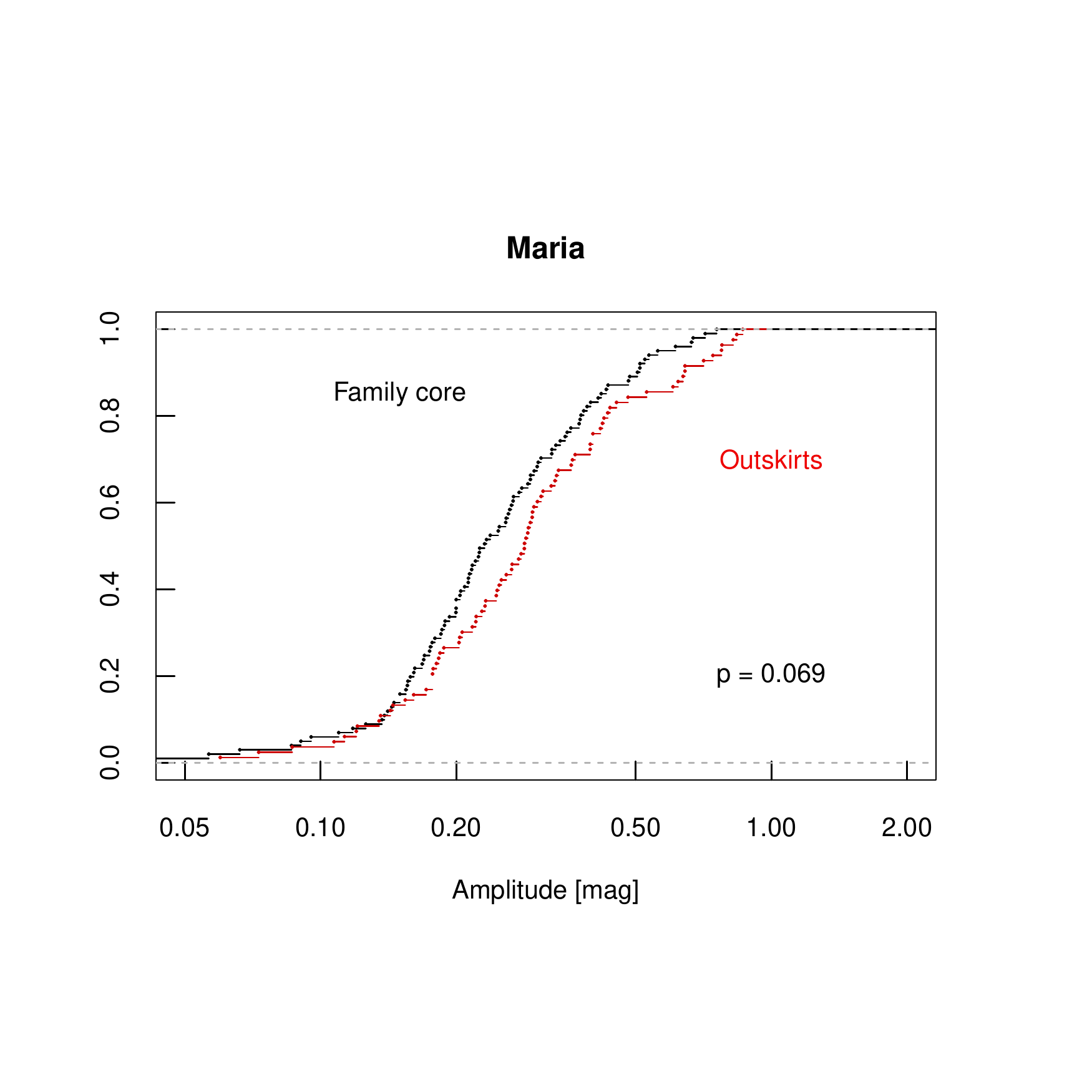}
\includegraphics[bb=7 50 473 420,clip,width=.33\textwidth]{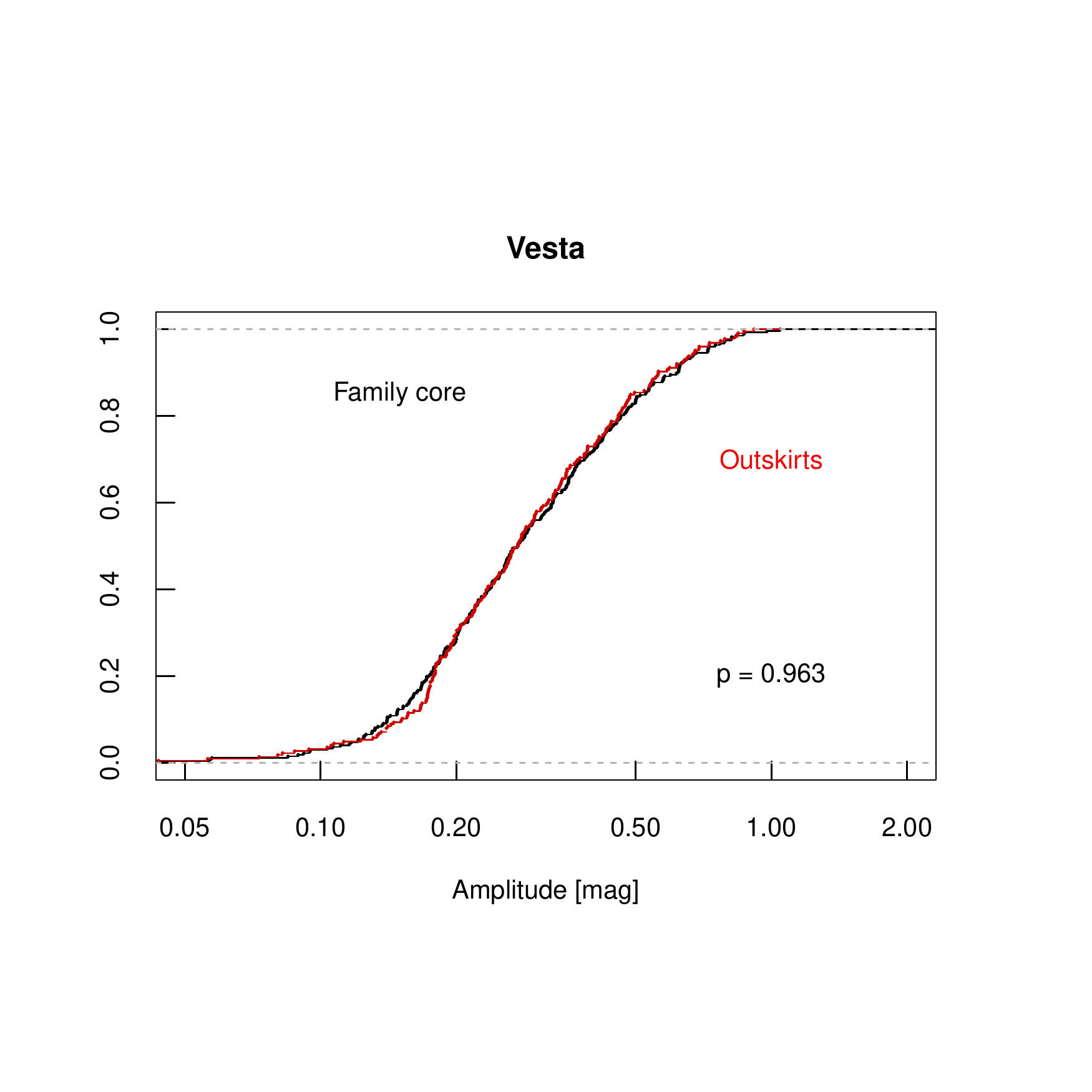}
}
\caption{Period and amplitude distributions in the cores and the outskirts of three families. }
\label{fig:cc}
\end{figure*}

We tried to detect differences between the youngest and oldest families due to a dependence on amplitude distribution and age. Therefore, we plotted the characteristic median amplitude in all examined asteroid families as a function of age (Fig. \ref{fig:yo_scatter}).

The asteroid families with dynamical age determinations are plotted with solid symbols. A log--linear fit confirms that the characteristic amplitude does indeed have an age dependence and it can be approximated by the following: 
\begin{equation}
Q(\rm {A},median) =  0.328 - 0.02067 \cdot {\rm age[Gyr]} \pm 0.025,
\label{formula}
\end{equation}
where the amplitude {\oldbf median} is expressed in magnitudes, and the confidence of this relation is $p=0.015$.

Since only an upper age estimate is available for several old families, we added them to the plot with an arrow indicating an upper estimate and with open symbols for less visual weight. These points are also consistent with the derived relation and also their distribution is consistent with an upper estimate for the age; however, more than half of them seem to be a rather good estimate. Family-only upper limits on their estimated ages were, however, omitted from the derivation of the formula from Eq. \ref{formula}.

\subsection{Cores versus outskirts}

There are known differences between the cores and outskirts of asteroid families, for example the size distribution and sometimes the colour distribution (such as the representation of different taxonomies) can be different \citep{2008Icar..198..138P}. The TSSYS-DR1 asteroid data set enabled us to reveal a possible similar structure of asteroid families in the period and amplitude distributions. 

The comparison was performed by dividing the population of the family into two parts, resulting in a lower number of data points in both subgroups of the comparison. This reduced the power of the tests, and because of this, only very stringent deviations could be detected when the number of test targets was small. Therefore we selected the five families with the most entries in TSSYS-DR1 for this specific comparison: the Vesta, Eos, Eunomia, Flora, and Maria families. The cores and outskirts were separated by calculating the Mahalanobis distance of the individual asteroids from the entire family. The split was set up so that 66.7\%{} of the members belonged to the core, and the remaining members belonged to the outskirts. The period and amplitude distributions were compared by following exactly the same methodology as in the case of comparing the families to each other. We show the results for three asteroid families in Fig. \ref{fig:cc}. 

Different rotation properties between the core and the outskirts were revealed in the case of Flora and Maria families. In both cases, the period distributions were found to be different at a larger significance, that is $1-p=99.6$ and $95.2$\%{}, respectively. In both families, the core members were found to rotate somewhat faster than the ones in the outskirts.

No internal structure of the amplitude distributions was detected in the Flora family, while in case of the Maria family we can suspect a slightly less median amplitude in the core. Due to the small number of the family members in DR1, the significance is low ($p=0.07$) and more data are needed to prove this hypothesis. Fortunately, later issues of the TESS asteroid database will contain the sufficient number of Maria (and other family) asteroids needed to repeat this analysis.

The Vesta family is an example where no internal differences were found in either the rotation period distribution or in that of the light curve amplitude. The two distributions are very close to each other. This is the largest asteroid family in our sample, and the lack of differences is so apparent that one can reasonably conclude on a high level of internal homogeneity in the Vesta family. The distributions in the Eos and Eunomia families are very similar to the case of Vesta, and we just mention the negative result here without including the figures.

We conclude that there are asteroid families which show a pattern in rotation properties, and there are other ones without a significant pattern. Differences in the period distribution are easier to detect than the variation of the amplitude distributions. These results again indicate that the evolution of rotation properties and shapes is a result of complex scenarios, which may have occurred differently at the level of the individual asteroid families.

\subsection{Principal component analysis}

The family-to-family variations of periods and amplitudes seemed to behave differently and emerged at a different level. We compared the complexity of this variation with a principal component analysis (PCA). The principal components are orthogonal axes with maximal variance of each; therefore, they give a set of linear combinations of sample parameters in such a way that almost the complete information about the data distribution is contained in a sub-space spanned by the first few principal components. The first component is usually a weighted mean of the entries, while the following components describe the evermore local deviations.

\begin{figure}
    \centering
    \includegraphics[viewport=4 100 472 366, width=\columnwidth]{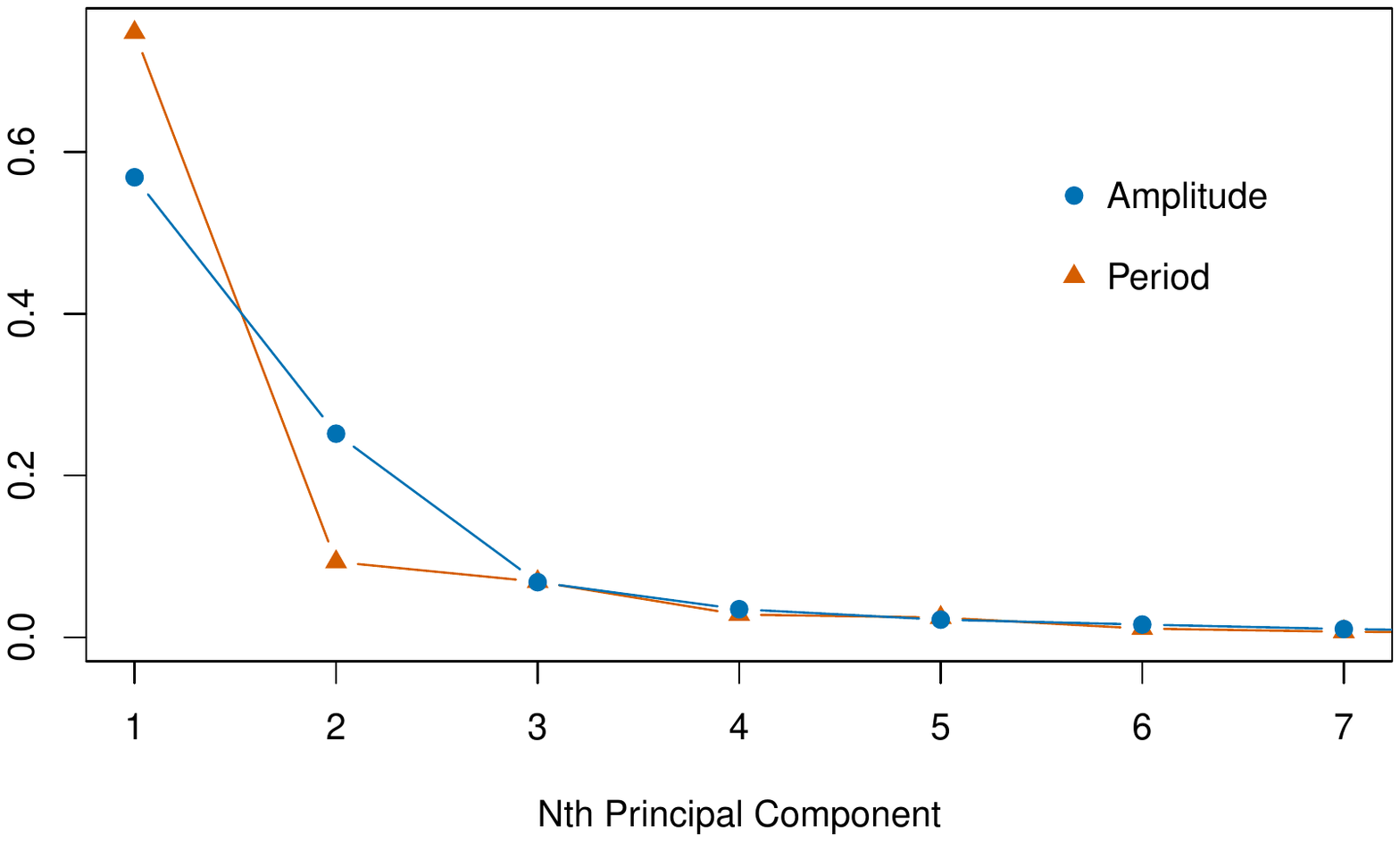}
    \vskip 0.4cm
    \caption{Explained variance of differences in amplitude and period distributions in families by the Nth principal components.}
    \label{fig:prcom}
\end{figure}

In Fig. \ref{fig:prcom} we show the explained variance by the consecutive principal components of the binned, non-cumulative, normalised amplitude and period distributions. This plot tells us that the period distributions follow a simpler pattern dominated by the amplitude of the first component, that is, the contribution of the usual pattern of the rotation period distribution in the family. There is little variation on the wings, likely reflecting the varying ratio of very slow and very fast rotators \citep[][]{2020ApJS..247...34S}, but these differences have to explain  only a small variance, below 10\%{}. On the other hand, the PCs of the amplitude distributions show a different pattern since 25\%{} of the variance is explained by the second principal component. This second-order component measures the ratio of low-amplitude asteroids to high-amplitude ones in the family, which has a significant variation between the families, and it is a major attribute of the different asteroid families.

\section{Summary}
\label{sec:summary}

In this paper, we have compared the rotation statistics of {\oldbf 16} asteroid families in the TSSYS-DR1 asteroid database and came to the following conclusions:

\begin{itemize}
    \item Fast rotating asteroids have a more {\oldbf spheroidal} shape than slower rotators. The most elongated asteroids are found in the 4--8 hour rotation period regime. Slowly  rotating asteroids, above an 11 hour period, have less of an elongated shape than asteroids with a 4--8 hour rotation period, but they are not as rounded as asteroids with a 1--3 hour period. 
    
    \item{} Some asteroid families including Hungaria, Maria, Juno, and Eos families  can be characterised by a unique amplitude and/or period distributions. However, more generally, the distribution of rotation periods and amplitudes are quite similar in the asteroid families. The period distribution can be characterised with a more significant family-to-family variation than the amplitude distributions.
    
    \item{} Young asteroid families can be characterised by a larger number of elongated bodies than old asteroid families. The shape distributions in the old families and the non-family asteroids are identical.
    
    \item{} Maria and Flora families have an internal structure in the rotation period distribution. Asteroids in the cores rotate faster than asteroids in the outskirts. The Vesta, Eos, and Eunomia families show an internal homogeneity of rotation properties.
    
    \item{} The family-to-family variation of asteroid shapes and the rotation period distributions is merely unrelated, indicating partly different scenarios behind the evolution of these properties.
    
\end{itemize} 

This comparison confirms again that the period distributions in asteroid families are more similar to each other than the amplitude distributions. This has important consequences regarding how the shape and the rotation properties evolve in the families. Because the amplitude distributions follow a more complex pattern, we suggest that the shape characteristics and the rotation rates evolve in at least partly different processes, and one major or several concurrent scenarios can shape the asteroids that have, at most, a slight effect on the distribution of rotation. Here, shaking from micro-impacts is a very good candidate since it continually evolves the asteroid to {\oldbf spheroidal} shapes, while the modification of the angular momentum is small and its direction is also somewhat stochastic.

\section*{Acknowledgments}
This paper includes data collected by the TESS mission. Funding for the TESS mission is provided by the NASA Explorer Program. This  project  has  been  supported  by  the  Hungarian National Research, Development and Innovation Office (NKFIH) grants GINOP-2.3.2-15-2016-00003, K-125015, K-138962 and KKP-137523, a PRODEX Experiment Agreement No. 4000137122, the Lend\"ulet LP2018-7/2021 grant of the Hungarian Academy of Sciences and the support of the city of Szombathely. Cs.~Kalup was supported by the \'UNKP-19-2, \'UNKP-20-2 and \'UNKP-21-2 New National Excellence Programs of the Ministry of Innovation and Technology from the source of the National Research, Development and Innovation Fund. Zs.B. acknowledges the support by the J\'anos Bolyai Research Scholarship of the Hungarian Academy of Sciences. The data presented in this paper were obtained from the Mikulski Archive for Space Telescopes (MAST). STScI is operated by the Association of Universities for Research in Astronomy, Inc., under NASA contract NAS5-26555. Support for MAST for non-HST data is provided by the NASA Office of Space Science via grant NNX09AF08G and by other grants and contracts. The authors thank the hospitality the Veszpr\'em Regional Centre of the Hungarian Academy of Sciences (MTA VEAB) where parts of this project were carried out.
We also would like to thank our anonymous referee for his/her help.

\bibliographystyle{aa}
\bibliography{references}
%\onecolumn

\begin{appendix}
\section{Light curve periods and amplitudes of asteroid families}

\begin{figure*}[!ht]
    \centering
    \includegraphics[viewport=124 242 495 781,width=14.5cm]{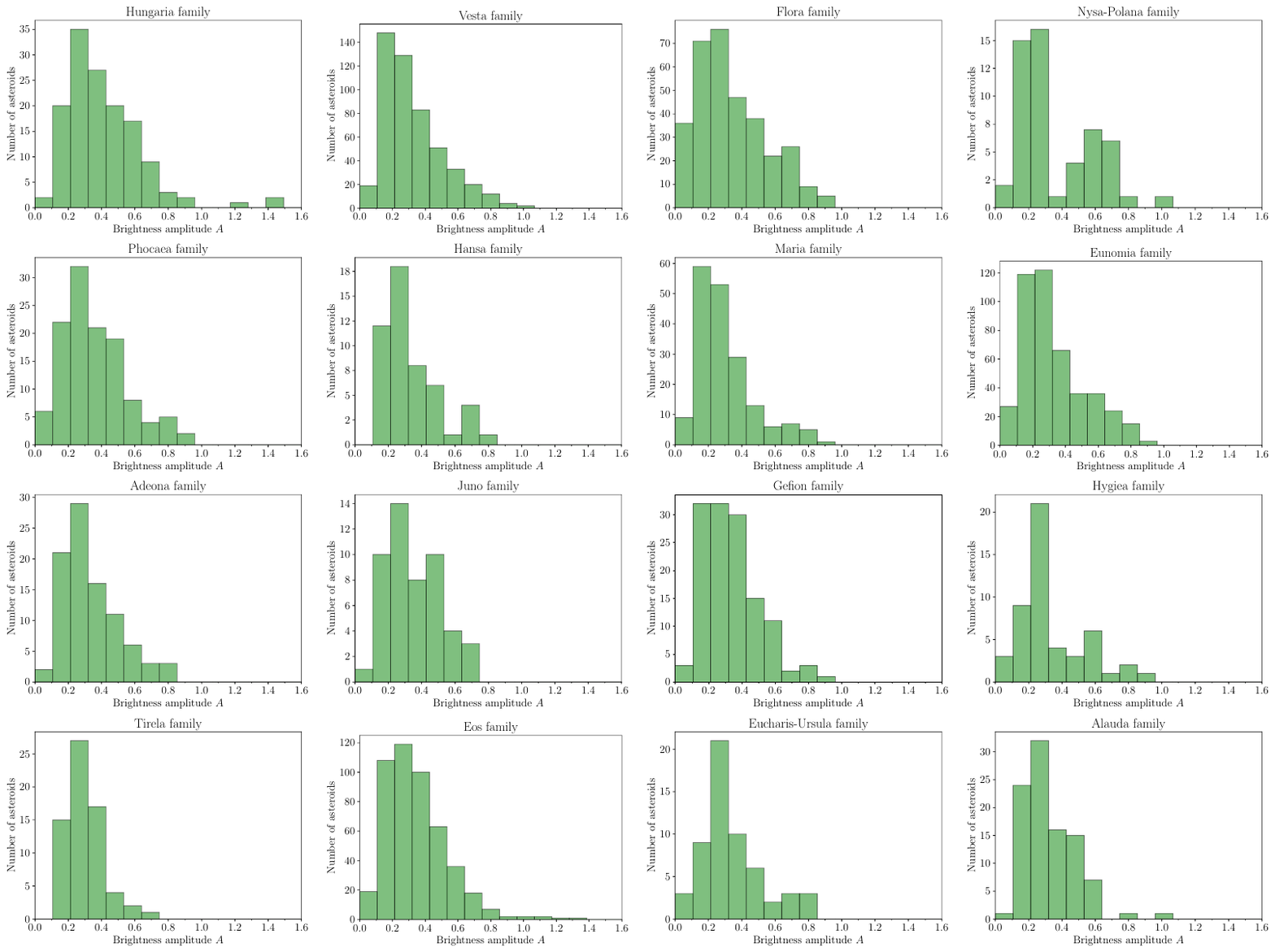}
    \caption{Distribution of the peak-to-peak light curve amplitude in the examined families, sorted by increasing mean solar distance.}
    \label{fig:ampfigures}
\end{figure*}

\begin{figure*}
    \centering
    \includegraphics[viewport=124 242 495 781,width=14.5cm]{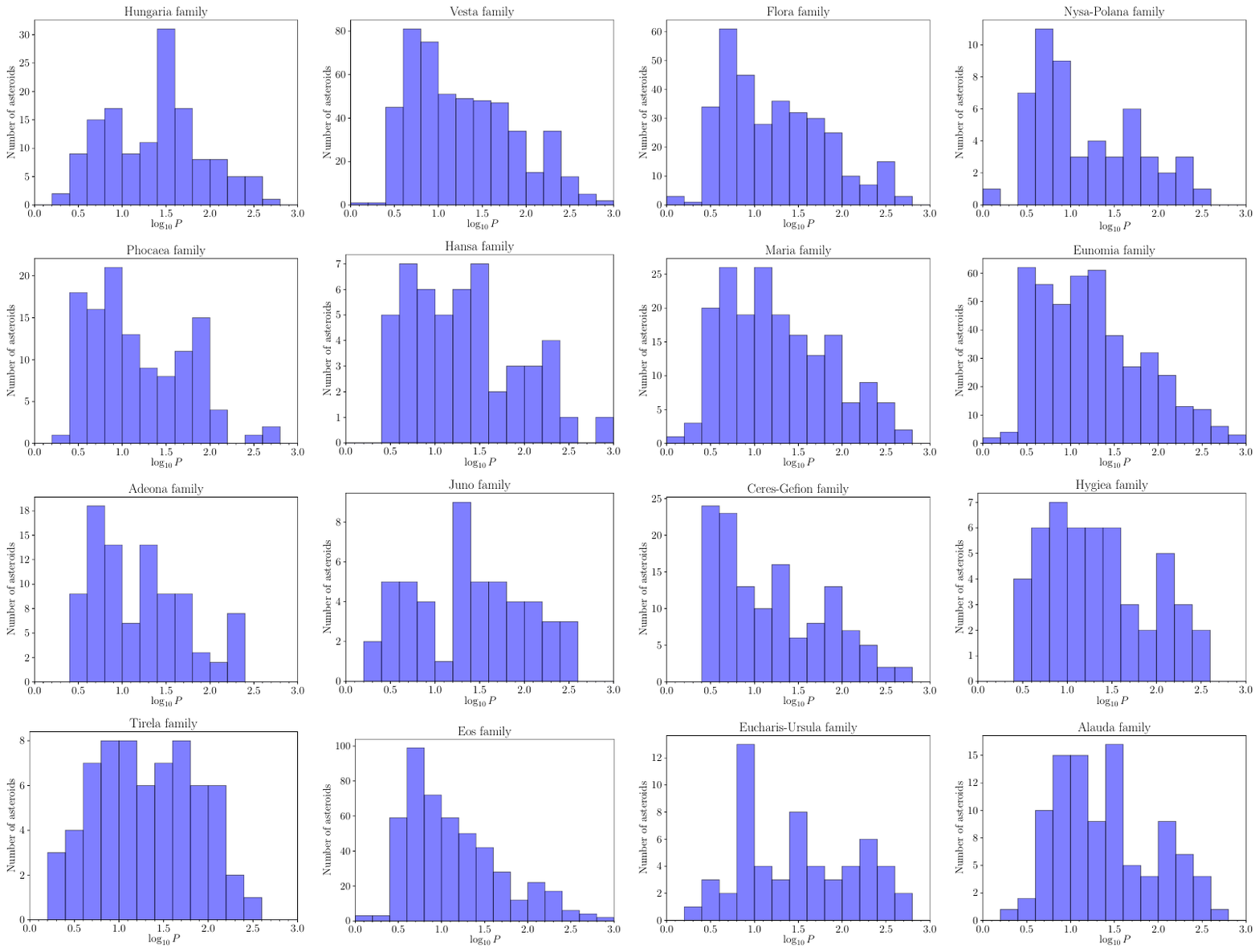}
    \caption{Same as Fig. \ref{fig:ampfigures}, but showing the distribution of rotation periods.}
    \label{fig:perfigures}
\end{figure*}

Below we present the distribution of light curve amplitudes (Fig.~\ref{fig:ampfigures}) and periods (Fig.~\ref{fig:perfigures}) of TESS light curves of asteroids in different families.

\end{appendix}

\end{document}